\documentclass[aps,prd,superscriptaddress,twocolumn,showpacs]{revtex4}
\usepackage{graphicx}
\usepackage{epstopdf}
\usepackage{amsmath}
\usepackage{amsfonts}
\usepackage{amssymb}
\usepackage{latexsym}
\usepackage{verbatim}
\begin{document}
\title{Lorentz-symmetry violating physics in a supersymmetric scenario in $(2+1)$-D}
\author{L. D. Bernal} \email{ldurand@cbpf.br}
\affiliation{Centro Brasileiro de Pesquisas F\'{i}sicas (CBPF), Rio de Janeiro, RJ, Brasil} \author{Patricio Gaete} \email{patricio.gaete@usm.cl} 
\affiliation{Departamento de F\'{i}sica and Centro Cient\'{i}fico-Tecnol\'ogico de Valpara\'{i}so-CCTVal, Universidad T\'{e}cnica Federico Santa Mar\'{i}a, Valpara\'{i}so, Chile}
\author{Y. P. M. Gomes} \email{ymuller@cbpf.br}
\affiliation{Centro Brasileiro de Pesquisas F\'{i}sicas (CBPF), Rio de Janeiro, RJ, Brasil} 
\author{Jos\'{e} A. Helay\"{e}l-Neto}\email{helayel@cbpf.br}
\affiliation{Centro Brasileiro de Pesquisas F\'{i}sicas (CBPF), Rio de Janeiro, RJ, Brasil} 
\date{\today}

\begin{abstract}
We study the dimensional reduction for a $(3+1)$-dimensional Lorentz Violating Lagrangian in the matter and gauge sectors in a Supersymmetric scenario. We thus obtain an effective model for photons induced by the effects of supersymmetry in our framework with Lorentz-symmetry violating. This model is discussed in connection with the attainment of a static potential between charged particles. Our calculation is done within the framework of the gauge-invariant, but path-dependent variable formalism. We thus find that the interaction energy displays a screening part, encoded by Bessel functions, and a linear confining potential.
\end{abstract}
\maketitle

\section{Introduction}

The Standard Model of Particle Physics (SM) provides a very satisfactory description of fundamental processes within a scale of a few hundred GeVs. Appealing TeV accessible energies, the LHC has been scrutinizing the SM searching for new physics which may stem from higher energies; it is the core of Supersymmetry (SUSY), quantum gravity effects and extra-dimensions.

The intensive activity on Lorentz Symmetry violation (LSV) encompasses broad variety of phenomena from ground-based experiments such as atomic physics, to accelerator physics and astrophysics. LSV takes place at vary high energies, close to the Planckian scale, but it may be felt at accessible regions of our observations and may point to new patches to understand more fundamental physics beyond the SM \cite{KPRL89,KosteleckyPRL91,Kostelecky89,KosteleckyPRD89,Potting91, Potting96,Potting95,DColladay97,Colladay98,Russell11,Bailey06,Chkareuli,Pospelov,Bailey04, Betschart,Tasson09,Lane,Lehnert,Mund:2019nap}.

It is reasonable to set up the discussion of LSV in a context where SUSY provides a good scenario. Our present understanding is that SUSY is broken at a lower scale, so that LSV takes place at a scale where SUSY should not be disregarded. In the works of Ref.\cite{MacD,Lehum,Belich:2013rma,Belich:2015qxa} we quote a number of papers where LSV is inspected in close connection with SUSY.

In the present contribution we pursue an investigation that connects LSV and SUSY no more in $(1+3)$-D, as it is currently done. We focus our attention on a $(1+2)$-dimensional space-time, so that planar phenomena may be contemplated. Though planar physics must directly interest Condensed Matter Phenomena and those, in principle, are not immediately concerned with Lorentz symmetry, we know that non-relativistic effects derived from Lorentz-invariant models are relevant for Condensed Matter Physics.

On the other hand, more recently, SUSY has appeared as an emergent symmetry in Condensed Matter systems \cite{Yu:2019opk,Gao:2017bqf,Rahmani:2015qpa,Grover:2013rc}. In view of that we wish, in this Letter, to report on a study we have undertaken which relates LSV, SUSY and $(1+2)$-D physics. We wish to understand how physical parameters associated to LSV and SUSY may interfere on planar models which might be of interest for describing low dimensional Condensed Matter physics.

In previous publications \cite{Belich:2013rma,Belich:2015qxa}, we have worked out effective photonic actions upon integrating over SUSY degrees of freedom, like the photino, for example. In the present Letter, we consider LSV in the matter sector (the specific model shall be present in the sequel) in four space-time dimensions in presence of SUSY. The supersymmetric LSV model will undergo supersymmetric dimensional reduction to $(1+2)$-dimensions and the reduced $3$D action will exhibit anisotropic parameters that are inherited from the $4$D mother model. 

An effective photonic action is considered to discuss the interaction energy for two probes charges. In other words, our purpose here is to further elaborate on the physical content of this new electrodynamics (photonics action) on a physical observable. To do this, we shall work out the static potential for this electrodynamics along the lines of \cite{Gaete:2018nwq,Gaete:2016tcn}. The advantage of using this development lies in the fact that the interaction energy between two static charges is obtained once a judicious identification of the physical degrees of freedom is made. As will be seen, this new electrodynamics is analogous to that encountered in $D=3$ models of antisymmetric tensor fields that emerges from the condensation of topological defects, as a consequence of the Julia -Thoulousse mechanism \cite{Gaete:2005ht}. This same potential profile is obtained in the case of condensation of charged scalars in $D=3$ dimensions \cite{Gaete:2009ev}. In other terms, in this work we are concerned with the physical content associated with a ``sort of duality'', where duality refers to an equivalence between two or more quantum field theories whose corresponding classical theories are different.

We would like to point out a number of interesting articles that discuss relevant physical aspects of Lorentz-symmetry violating models in $(1+2)$ dimensions, both without \cite{Ferreira:2004hx,Ferreira:2004ax,Casana:2011vh,Casana:2011du,Casana:2011fe,Ferreira:2019jbx,Casana:2015kna} and with \cite{Zarro,Petrov,Farias,Silva} SUSY taken into account.

Our work is organized to the following outline: in Section II, we perform the dimensional reduction for a $(3+1)$-dimensional LV-Lagrangian in the matter and gauge sectors along the lines of \cite{Scherk}. In Section II, we compute the interaction energy for a fermion-antifermion pair in this new electrodynamics. Interestingly enough, for this new electrodynamics, the static potential profile contains a linear term, leading to the confinement of static charges. Finally, some concluding remarks are presented in Sect. IV. An Appendix follows where we cast our conventions and notations.

In our conventions the signature of the metric is $(+1, -1, -1)$.
 
\section{SUSY matter Lorentz-breaking Lagrangian}

\subsection{Matter Lagrangian}

We begin our discussion with the fermionic Lagrangian. We first observe that
in $(1+3)$-D the LV-Lagrangian for the matter field $\Psi \ $ reads
\begin{equation}
\mathcal{L}_f = \bar{\Psi} \Gamma^{\hat{\mu}}(i \partial_{\hat{\mu}} + \bar{a}_{\hat{\mu}} +\bar{b}_{\hat{\mu}}\Gamma_5 - m) \Psi, \label{susy3d05}
\end{equation}
where $\bar{b}^{\hat{\mu}} = \frac{1}{4}(W+V)^{\hat{\mu}} - b^{\hat{\mu}}$ , $\bar{a}^{\hat{\mu}} = \frac{1}{4}(W-V)^{\hat{\mu}} - a^{\hat{\mu}}$ , $W^{\hat{\mu}}(V) = \bar{\Lambda}_{+(-)}
\Gamma^{\hat{\mu}} \Gamma_5 \Lambda_{+(-)}$ are given in Ref.\cite{}  and represent the LV-backgroung parameters.  Besides, the bosonic part is given by
\begin{eqnarray}\nonumber
&&\mathcal{L}_b = -\frac{1}{2} \phi_1^* \Big{(} \partial^{\hat{\mu}} \partial_{\hat{\mu}} + m^2 + i 2 \sqrt{2}(a+b)^{\hat{\mu}} \partial_{\hat{\mu}} \Big{)} \phi_1 \\ &&-\frac{1}{2} \phi_2^* \Big{(}\partial^{\hat{\mu}} \partial_{\hat{\mu}} + m^2 + i 2 \sqrt{2}(a+b)^{\hat{\mu}} \partial_{\hat{\mu}} \Big{)} \phi_2. \label{susy3d10}
\end{eqnarray}

It should be further noted that beyond the quadratic terms, this method brings us a different kind of interaction between scalar and fermionic fields through the following interaction Lagrangian
\begin{equation}
\mathcal{L}_{int} = \Phi^\dagger v + h.c., \label{susy3d15}
\end{equation}
where $\Phi^\dagger = \begin{pmatrix}
\phi_1^* & \phi_2^*
\end{pmatrix}$ and $v = (i \bar{\Lambda} \Gamma \cdot\partial +\frac{m}{2} \bar{\Lambda}_R) \begin{pmatrix}\Psi\\  \Gamma_5 \Psi\end{pmatrix} $ . 

We can now to apply the dimensional reduction in the total lagrangian $\mathcal{L}_{tot} = \mathcal{L}_f + \mathcal{L}_b + \mathcal{L}_{int} $.

Following the conventions given in the Appendix, the corresponding dimensional reduction of the Lagrangian is carried out with $\Psi = \begin{pmatrix}\Psi_1 & \Psi_2 \end{pmatrix}^T$. We thus obtain
\begin{eqnarray}
\mathcal{L}_f \mspace{-6mu} &=&\mspace{-8mu}  \begin{pmatrix}\bar{\Psi}_1 & \bar{\Psi}_2\end{pmatrix} \nonumber\\
&\times&\mspace{-8mu} \begin{pmatrix}i \gamma \cdot \partial + \gamma\cdot\bar{a}- \bar{b}_3 + m  & - i \gamma \cdot \bar{b} + i \bar{a}_3 \\- i \gamma \cdot \bar{b} + i \bar{a}_3 & -i \gamma \cdot \partial - \gamma\cdot\bar{a}+ \bar{b}_3 + m\end{pmatrix} \nonumber\\
&\times& \mspace{-8mu}\begin{pmatrix}\Psi_1 \\ \Psi_2 \end{pmatrix}, \label{susy3d20}
\end{eqnarray}
where $\bar{\Psi}_1 = \Psi_1^{\dagger} \gamma^0 $. The bosonic and the mixed part can be evaluated together and they are given by
\begin{equation}
\mathcal{L}_{b+int} = \Phi^{\dagger} O(\partial) \Phi + \Phi^{\dagger} v + v^{\dagger} \Phi, \label{susy3d25}
\end{equation}
where, 
\begin{equation}
O(\partial) = \frac{1}{2} \Big(\partial^{\mu} \partial_{\mu} + m^2 + i 2 \sqrt{2}(a+b)^{\mu} \partial_{\mu}\Big), \label{susy3d30}
\end{equation} 
and 
\begin{eqnarray}
v&=&(i \bar{\Lambda}_1 \gamma \cdot\partial +\frac{m}{2} \bar{\Lambda}_{1R}) \begin{pmatrix}1 & 0\\0 & 1 \end{pmatrix} \begin{pmatrix}\Psi_1 \\ \Psi_2 \end{pmatrix} \nonumber\\
&+& (i \bar{\Lambda}_2 \gamma \cdot\partial +\frac{m}{2} \bar{\Lambda}_{2R})\begin{pmatrix}0 & 1 \\ -1 & 0 \end{pmatrix} \begin{pmatrix}\Psi_1 \\ \Psi_2 \end{pmatrix} \nonumber\\
&=& \begin{pmatrix} \bar{Q}_1 & \bar{Q}_2 \\ - \bar{Q}_2 & \bar{Q}_1 \end{pmatrix} \begin{pmatrix}\Psi_1 \\ \Psi_2 \end{pmatrix}. \label{susy3d35}
\end{eqnarray} 

The $\bar{Q}_1$ and $\bar{Q}_2$ terms are given by $\bar{Q}_1=(i \bar{\Lambda}_1 \gamma \cdot\partial +\frac{m}{2} \bar{\Lambda}_{1R}) $ and $\bar{Q}_2 = (i \bar{\Lambda}_2 \gamma \cdot\partial +\frac{m}{2} \bar{\Lambda}_{2R})$. 

It is worth noting here that by making use of the shift in the field $\Phi \rightarrow \Phi + O(\partial)^{-1}v$, the previous equation can be written alternatively in the form 
\begin{equation}
\mathcal{L}_{b+ int} = \Phi^{\dagger} O(\partial) \Phi - v^{\dagger} O(\partial)^{-1} v, \label{susy3d40}
\end{equation} 
where 
\begin{eqnarray}
v^{\dagger} O(\partial)^{-1} v \mspace{-6mu} &=&\mspace{-8mu} \begin{pmatrix}\bar{\Psi}_1 & \bar{\Psi}_2\end{pmatrix} \nonumber\\
&\times& \mspace{-10mu}  \begin{pmatrix} Q_1 \bar{Q_1} - Q_2 \bar{Q_2} &Q_1 \bar{Q_2} + Q_2 \bar{Q_1}\\ -Q_1 \bar{Q_2} - Q_2 \bar{Q_1} & Q_2 \bar{Q_2} -  Q_1 \bar{Q_1} \end{pmatrix} \nonumber\\
&\times&\mspace{-8mu} O^{-1}(\partial) \begin{pmatrix}\Psi_1 \\ \Psi_2 \end{pmatrix},  \label{susy3d45}
\end{eqnarray}
and $O^{-1}(\partial) = \Big(\partial^{\mu} \partial_{\mu} + m^2 + i 2 \sqrt{2}(a+b)^{\mu} \partial_{\mu}\Big)^{-1}$. 

We thus find that the total Lagrangian becomes
\begin{eqnarray}
\mathcal{L}_{tot}\mspace{-6mu} &=& \mspace{-8mu} \begin{pmatrix}\bar{\Psi}_1 & \bar{\Psi}_2\end{pmatrix} \nonumber\\
&\times& \mspace{-8mu}  \begin{pmatrix}i \gamma \cdot \partial + \gamma\cdot\bar{a}- \bar{b}_3 + m  & - i \gamma \cdot \bar{b} + i \bar{a}_3 \\- i \gamma \cdot \bar{b} + i \bar{a}_3 & -i \gamma \cdot \partial - \gamma\cdot\bar{a}+ \bar{b}_3 + m\end{pmatrix}\nonumber\\
&\times&\mspace{-8mu} \begin{pmatrix}\Psi_1 \\ \Psi_2 \end{pmatrix}  \nonumber\\
&+&\mspace{-8mu}  \sum_{i=1}^2 \phi_i O(\partial) \phi_i + \begin{pmatrix}\bar{\Psi}_1 & \bar{\Psi}_2\end{pmatrix} \Omega(\partial, \Lambda) O^{-1}(\partial) \begin{pmatrix}\Psi_1 \\ \Psi_2 \end{pmatrix}, \nonumber\\
\label{susy3d50}
\end{eqnarray}
where $ \Omega(\partial, \Lambda) =  \begin{pmatrix} Q_1 \bar{Q_1} - Q_2 \bar{Q_2} &Q_1 \bar{Q_2} + Q_2 \bar{Q_1}\\ -Q_1 \bar{Q_2} - Q_2 \bar{Q_1} & Q_2 \bar{Q_2} -  Q_1 \bar{Q_1} \end{pmatrix}$. 

We now want to extend what we have done to the gauge sector.

\subsection{CPT-odd Gauge Lagrangian}

By the introduction of a background scalar superfield 
\begin{eqnarray}
S &=& s + \sqrt2 \theta \chi + i\bar{\theta}\sigma^{\hat{\mu}} \theta \partial_{\hat{\mu}} s + \theta^2 F + \frac{i}{\sqrt2}\theta^2 \bar{\theta}\bar{\sigma}^{\hat{\mu}} \partial_{\hat{\mu}}\chi \nonumber\\
&-& \frac{1}{4} \bar{\theta}^2 \theta^2 \Delta \ s, \label{susy3d55}
\end{eqnarray}
with the properties $(s + s^*) = 0$, $ (s - s^*) = - \frac{i}{2} x_{\hat{\mu}} \xi^{\hat{\mu}}$ and $\partial_{\hat{\mu}} \chi = 0$, we are able to write a LV-action. These properties have a meaning that the SUSY breaks down and generates a non-null vector background $\xi$ and a non-null fermionic parameter $\chi$. The action is given by:
\begin{equation}
S_{CPT-odd} = \int d^4 x d^4 \theta \Big{(} W^\alpha (D_\alpha V) S + W^{\dot{\alpha}} (D_{\dot{\alpha}} V) S \Big{)}, \label{susy3d60}
\end{equation}
where $V$ is the Vector superfield in the Wess-Zumino gauge and $W^\alpha = -\frac{1}{4}(\bar{D})^2D^{\alpha} V $. Rewriting in terms of the component fields the total Lagrangian is written as, $\mathcal{L}_{tot-gauge} = \mathcal{L}_{A} + \mathcal{L}_{ph} + \mathcal{L}_{int-gauge}$, where 
\begin{equation}
\mathcal{L}_A = -\frac{1}{4}{F_{\hat{\mu} \hat{\nu}} }^2 + \frac{1}{2} \epsilon^{\hat{\mu} \hat{\nu} \hat{\alpha} \hat{\beta}} \xi_{\hat{\mu}} A_{\hat{\nu}} F_{\hat{\alpha} \hat{\beta}}. \label{susy3d65}
\end{equation}
For the photino ($\lambda$) we have (with fermionic Lorentz breaking parameter $\chi$)
\begin{eqnarray}
\mathcal{L}_{ph} &=& - \frac{i}{2} \bar{\lambda}\Gamma^{\hat{\mu}}\partial_{\hat{\mu}} \lambda + \phi \bar{\lambda} \lambda \nonumber \\
&-&i \rho \bar{\lambda} \Gamma_5 \lambda - \bar{V}_{\hat{\mu}} \bar{\lambda} \Gamma^{\hat{\mu}}\Gamma_5 \lambda, \label{susy3d70}
\end{eqnarray}
with $\phi = \Big[Re(F)+ \frac{1}{4}\bar{\chi} \chi \Big]$ , $\rho = \Big[Im(F) + \frac{i}{4}\bar{\chi}\Gamma_5 \chi \Big]$ and $\bar{V}_{\hat{\mu}} = \frac{1}{4}\Big[V_{\hat{\mu}}+ \bar{\chi}\Gamma_{\hat{\mu}}\Gamma_5 \chi \Big]$. This procedure also given a new interaction term between the photon and the photino field, and this term is given by
\begin{equation}
\mathcal{L}_{int-gauge} = \sqrt{2} \bar{\lambda}\Gamma^{\hat{\mu} \hat{\nu}} \Gamma_5 \chi F_{\hat{\mu} \hat{\nu}}. \label{susy3d75}
\end{equation}

Now, we apply the dimensional reduction to the Lagrangean $\mathcal{L}_{tot-gauge} = \mathcal{L}_A + \mathcal{L}_{ph} + \mathcal{L}_{int-gauge}$.

From the dimensional reduction scheme of the Appendix, we write $A^{\hat{\mu}} = (A^\mu, \varphi)$ , 
$\lambda = \begin{pmatrix}\lambda_1 & \lambda_2 \end{pmatrix}^T$ and $\chi = \begin{pmatrix}\chi_1  & \chi_2 \end{pmatrix}^T$, we have
\begin{eqnarray}
\mathcal{L}_A &=& -\frac{1}{4} F_{\mu \nu}^2 + \frac{1}{2}\partial_\mu \varphi \partial^\mu \varphi  -\frac{\varphi}{4}\varepsilon^{\mu \nu \alpha} \xi_\mu \partial_\nu A_\alpha \nonumber\\
&-&\frac{1}{2}\xi_3 \varepsilon^{\mu \nu \alpha}A_\mu \partial_\nu A_\alpha. \label{susy3d80}
\end{eqnarray}
The photino sector is given by
\begin{eqnarray}
\mathcal{L}_{ph} \mspace{-6mu}&=& \mspace{-8mu}\begin{pmatrix}\bar{\lambda}_1 & \bar{\lambda}_2 \end{pmatrix}\begin{pmatrix}-\frac{i}{2} \gamma \cdot \partial +  \phi - \bar{V}_3 &  \rho - i \gamma \cdot \bar{V} \\ -\rho - i \gamma \cdot \bar{V} & \frac{i}{2} \gamma \cdot \partial + \phi + \bar{V}_3 \end{pmatrix} \nonumber\\
&\times&\mspace{-8mu}\begin{pmatrix}\lambda_1 \\ \lambda_2 \end{pmatrix}, \label{susy3d85}
\end{eqnarray}
where $\bar{V}_{\mu} = \frac{1}{4}\Big[V_{\mu}+ \bar{\chi}\Gamma_{\mu}\Gamma_5 \chi \Big]$ and $\bar{V}_{\hat{\mu}} = \frac{1}{4}\Big[V_{\hat{\mu}}+ \bar{\chi}\Gamma_{\hat{\mu}}\Gamma_5 \chi \Big]$. The mixing terms can be rewritten as follows:
\begin{eqnarray}
\mathcal{L}_{int-gauge} &= &\sqrt{2} i (\bar{\lambda}_1\gamma^{\mu \nu} \chi_2 + \bar{\lambda}_2\gamma^{\mu \nu} \chi_1)F_{\mu \nu} \nonumber\\
&+& \sqrt{2}(\bar{\lambda}_1\gamma^{\mu} \chi_2 + \bar{\lambda}_2\gamma^{\mu} \chi_1) \partial_\mu \varphi. \label{susy3d90}
\end{eqnarray}
In a short way, we can rewrite the above equation as $\mathcal{L}_{int} = \bar{\Theta} \Upsilon $ , where $\bar{\Theta} = \begin{pmatrix}\bar{\lambda}_1 & \bar{\lambda}_2 \end{pmatrix}$ and
\begin{equation}
\Upsilon =\sqrt{2} \begin{pmatrix}\gamma \cdot F & \gamma \cdot \partial \varphi \\ \gamma \cdot \partial \varphi & \gamma \cdot F \end{pmatrix} \begin{pmatrix} \chi_2 \\ \chi_1 \end{pmatrix}. \label{susy3d95}
\end{equation}
Here $\gamma \cdot F = \gamma^{\mu \nu} F_{\mu \nu}$ and $\gamma \cdot \partial \varphi = \gamma^\mu \partial_\mu \varphi$. Manipulating the equation above and applying the same kind of shift used in the matter sector, we can rewrite the photino and mixing terms as
\begin{eqnarray}
\mathcal{L}_{ph+ int} &=& \bar{\Theta} O(\partial) \Theta + \bar{\Theta} \Upsilon = \bar{\Theta} O(\partial)\Big(\Theta +\frac{1}{2} O^{-1}(\partial) \Upsilon \Big) \nonumber\\
&+& \frac{1}{2} \bar{\Theta} \Upsilon, \label{susy3d100}
\end{eqnarray}
where $\tilde{O}(\partial) = \begin{pmatrix}-\frac{i}{2} \gamma \cdot \partial +  \phi - \bar{V}_3 &  \rho - i \gamma \cdot \bar{V} \\ -\rho - i \gamma \cdot \bar{V} & \frac{i}{2} \gamma \cdot \partial + \phi + \bar{V}_3 \end{pmatrix}$. 

With the shift $\Theta \rightarrow \Theta + \frac{1}{2} \tilde{O}^{-1}(\partial) \Upsilon $ we have, finally
\begin{equation}
\mathcal{L}_{ph+ int} = \bar{\Theta} \tilde{O}(\partial) \Theta - \frac{1}{4} \bar{\Upsilon} \tilde{O}^{-1} \Upsilon. \label{susy3d105}
\end{equation}
Thus, the final action of the CPT-odd gauge sector will be given by
\begin{eqnarray}
 \mathcal{L}_{tot-gauge} &=& -\frac{1}{4} F_{\mu \nu}^2 + \frac{1}{2}\partial_\mu \varphi \partial^\mu \varphi  -\frac{\varphi}{4}\varepsilon^{\mu \nu \alpha} \xi_\mu \partial_\nu A_\alpha  \nonumber\\
 &-&\frac{1}{2}\xi_3 \varepsilon^{\mu \nu \alpha}A_\mu \partial_\nu A_\alpha  \nonumber\\
 &-& \frac{1}{4} \bar{\Upsilon} \tilde O^{-1} \Upsilon +  \begin{pmatrix}\bar{\lambda}_1 & \bar{\lambda}_2 \end{pmatrix} \tilde{O}(\partial) \begin{pmatrix}\lambda_1 \\ \lambda_2 \end{pmatrix}.  \label{susy3d105b}
 \end{eqnarray}

In summary then, we have obtained a complete $(2+1)$-dimensional Lagrangian, which defines a new electrodynamics. In the following Section we compute the interaction energy between static point-like sources for this new electrodynamics. Following the same steps as the ones presented in our previous works \cite{Belich:2013rma,Belich:2015qxa}, we get an effective photonics-scalar Lagrangian which shall be the matter in the coming Section.

\section{Interaction energy}

As we have already expressed before, we now proceed to calculate the interaction energy between static point-like sources for the model under consideration by using the gauge-invariant but path-dependent variables formalism examine the interaction energy, along the lines of Ref.\cite{ Gaete:2018nwq,Gaete:2016tcn}. In this case the corresponding theory is governed by the Lagrangian density
\begin{eqnarray}
{\cal L} &=&  - \frac{1}{4}F_{\mu \nu }^2 + \frac{m}{4}{\varepsilon ^{\mu \nu \kappa }}{A_\mu }{F_{\nu \kappa }} + m{\varepsilon ^{\mu \nu \kappa }}{v_\mu }{F_{\nu \kappa }}\varphi  \nonumber\\
&+& \frac{1}{2}{\left( {{\partial _\mu }\varphi } \right)^2} + {t_{\mu \nu }}{F^{\mu \lambda }}F_\lambda ^\nu  + \alpha {t_{\mu \nu }}{F^{\mu \lambda }}\frac{\Delta }{{\bar \Delta }}F_\lambda ^\nu 
 \notag \\
&+& \beta {t_{\rho \lambda }}{F^{\mu \lambda }}\frac{{{\partial _\mu }{\partial _\nu }}}{{\bar \Delta }}{F^{\nu \rho }} + {s^\mu }{F_{\mu \nu }}{\partial ^\nu }\varphi  + {s_\lambda }{F^{\mu \lambda }}\frac{{{\partial _\mu }\Delta }}{{\bar \Delta }}\varphi, \nonumber\\
\label{susyBI-05}
\end{eqnarray}
where $\Delta \equiv \partial _\mu \partial ^\mu$, ${v_\mu } \equiv {\xi _\mu }$, ${t_{\mu \nu }}$ and $s_{\mu}$ are given in Section of the work in Ref. \cite{}. It is also important to observe that when we carry out the integration over the $\varphi$-field, we find the following effective theory:
\begin{eqnarray}
{\cal L}\mspace{-6mu} &=&\mspace{-8mu} -\frac{1}{4}{F_{\mu \nu }}\left( {1 - \frac{{4{m^2}{v^2}}}{\Delta }} \right){F^{\mu \nu }} + \frac{m}{4}{\varepsilon ^{\mu \nu \kappa }}{A_\mu }{F_{\nu \kappa }} \nonumber\\
&+&\mspace{-8mu} {t_{\mu \nu }}{F^{\mu \lambda }}F_\lambda ^\nu  + \alpha {t_{\mu \nu }}{F^{\mu \lambda }}\frac{\Delta }{{\bar \Delta }}F_\lambda ^\nu  + 
\beta {t_{\rho \lambda }}{F^{\mu \lambda }}\frac{{{\partial _\mu }{\partial _\nu }}}{{\bar \Delta }}{F^{\nu \rho }}  \notag \\
&+&\mspace{-8mu} 2{v_\mu }{v_\nu }{F^{\mu \lambda }}\frac{{{m^2}}}{\Delta }F_\lambda ^\nu  - m{\varepsilon _{\rho \xi \sigma }}{v^\xi }{s_\lambda }{F^{\mu \lambda }}\left( {\frac{1}{{\bar \Delta }} - \frac{1}{\Delta }} \right){\partial _\mu }{F^{\rho \sigma }} \nonumber\\
&+&\mspace{-8mu}\frac{1}{2}{s_\lambda }{s_\nu }{F^{\mu \lambda }}\left( {\frac{\Delta }{{{{\left( {\bar \Delta } \right)}^2}}} - \frac{2}{{\bar \Delta }} + \frac{1}{\Delta }} \right){\partial _\mu }{\partial _\rho }{F^{\nu \rho }}. \label{susyBI-10}
\end{eqnarray}
However, as was mentioned before, this paper is aimed at studying the static potential of the above
theory, a consequence of this is that one may replace $\Delta$ by $-\nabla ^2
$ in Eq.(\ref{susyBI-10}). Furthermore, we recall that the only
non-vanishing $t_{\mu \nu }$-terms are the diagonal ones, since, as already
anticipated, $t_{\mu \nu }$ can be brought into a diagonal form. 
Without loss of generality, we may always choose $t_{00}\ne 0$.
Restricting our considerations to the ${v^i} \ne 0$ and ${v^{ij}} = 0$ ($v_{0}=0$) case (referred to as the space-like background in what follows), the effective Lagrangian becomes
\begin{eqnarray}
{\cal L}\mspace{-6mu} &=&\mspace{-8mu} - \frac{1}{4}{F_{\mu \nu }}\left( {1 - \frac{{4{m^2}{{\bf v}^2}}}{{{\nabla ^2}}}} \right){F^{\mu \nu }} + \frac{m}{4}{\varepsilon ^{\mu \nu \kappa }}{A_\mu }{F_{\nu \kappa }} \nonumber\\
&+&\mspace{-8mu} {t_{00}}\frac{{\left( {{A_2} + \alpha } \right)}}{{{A_2}}}{F^{i0}}{\cal O}{F^{i0}} + 
 - \beta \frac{{{t_{00}}}}{{{A_2}}}{F^{i0}}\frac{{{\partial _i}{\partial _j}}}{{\left( {{\nabla ^2} - {X^2}} \right)}}{F^{j0}} \notag \\
&+&\mspace{-8mu} \frac{{2m}}{{{A_2}}}\left( {{\bf v} \cdot {\bf s}} \right){F^{j0}}{{\cal O}^ \prime }{\partial _j}B + B{{\cal O}^{ \prime  \prime }}B -A_{0}J^{0}, \label{susyBI-15}
\end{eqnarray}
where $B$ is the magnetic field ($B = {\varepsilon _{ij}}{\partial ^i}{A^j}$), ${A_1} = {\mu ^2}$ and ${A_2} \equiv \left( {{c^{ii}} - 1} \right) = \left( {k - 1} \right)$. Notice that these $A_{1}$ and $A_{2}$ are not to be confused with the components of the photon field. Nevertheless ${\cal O} \equiv \left[ {\frac{{{\nabla ^4} - p{\nabla ^2} - q}}{{{\nabla ^2}\left( {{\nabla ^2} - {X^2}} \right)}}} \right]$, 
${{\cal O}^ \prime } \equiv \left[ {\frac{{\left( {1 - {A_2}} \right){\nabla ^2} + {A_1}}}{{{\nabla ^2}\left( {{\nabla ^2} - {X^2}} \right)}}} \right]$ and ${{\cal O}^{ \prime  \prime }} \equiv \left[ {\frac{{- {\bar v}{\nabla ^2} + w}}{{{\nabla ^2}\left( {{\nabla ^2} - {X^2}} \right)}}} \right]$. Here $p = \frac{{\left( {{t_{00}}{A_1} - 2{m^2}{{\bf v}^2}{A_2}} \right)}}{{{t_{00}}\left( {{A_2} + \alpha } \right)}}$, $q = \frac{{\left( {2{m^2}{{\bf v}^2}{A_1}} \right)}}{{{t_{00}}\left( {{A_2} + \alpha } \right)}}$, ${X^2} = \frac{{{A_1}}}{{{A_2}}}$, $\bar v = 4{m^2}{{\bf v}^2}$ and $w = 4{m^2}{{\bf v}^2}{X^2}$.

Having characterized the model under study, we shall now examine the interaction energy. To this end, we shall first consider the Hamiltonian framework for this model. We thus find that the canonical momenta are found to be 
${\Pi ^\mu } = \left( {1 - \frac{{4{m^2}{{\bf v}^2}}}{{{\nabla ^2}}}} \right){F^{\mu 0}} + \frac{m}{2}{\varepsilon ^{\mu 0\lambda }}{A_\lambda } + 2{t_{00}}\frac{{\left( {{A_2} + \alpha } \right)}}{{{A_2}}}{F^{\mu 0}} + \frac{{2\beta {t_{00}}}}{{{A_2}}}\frac{{{\partial ^\mu }{\partial _i}}}{{\left( {{\nabla ^2} - {X^2}} \right)}}{F^{i0}} - \frac{{4m}}{{{A_2}}}\left( {{\bf v} \cdot {\bf s}} \right){O^ \prime }{\partial ^\mu }B$. From this expression it follows that $\Pi^0=0$, which is the usual primary constraint equation. It should be further noted that the remaining non-zero momenta are, ${\Pi ^i} = \left\{ {\left( {1 - \frac{{4{m^2}{{\bf v}^2}}}{{{\nabla ^2}}}+2{t_{00}}\frac{{\left( {{A_2} + \alpha } \right)}}{{{A_2}}}{\cal O}} \right){\delta ^{ij}} - \frac{{2\beta {t_{00}}}}{{{A_2}}}\frac{{{\partial ^i}{\partial ^j}}}{{\left( {{\nabla ^2} - {X^2}} \right)}}} \right\}{F^{j0}} + \frac{m}{2}{\varepsilon ^{ij}}{A_j} - \frac{{4m}}{{{A_2}}}\left( {{\bf v} \cdot {\bf s}} \right){{\cal O}^ \prime }{\partial ^i}B$.

We thus find that the canonical Hamiltonian takes the form 
\begin{eqnarray}
{H_C} \mspace{-6mu}&=&\mspace{-8mu} \int {{d^2}x} \left\{ { - {A_0}\left( {{\partial _i}{\Pi ^i} + \frac{m}{2}{\varepsilon ^{ij}}{\partial _i}{A_j} - {J^0}} \right)} \right\} \nonumber\\
 &+&\mspace{-8mu} \int {{d^2}x} \left\{ {\frac{1}{2}{E_i}\Lambda {D_{ij}}{E_j} + \frac{{2m}}{{{A_2}}}\left( {{\bf v} \cdot {\bf s}} \right){E^i}{O^\prime}{\partial _i}B} \right\}\nonumber\\
 &+&\mspace{-8mu} \int {{d^2}x} \left\{ {\frac{1}{2}B\left( {1 - \frac{{4{m^2}{v^2}}}{{{\nabla ^2}}} - 2{O^{\prime \prime}}} \right)B}\right\}, 
\label{susyBI-20}
\end{eqnarray}
where $\Lambda  = \frac{{\left[ {{\nabla ^4} - a{\nabla ^2} + b} \right]}}{{{\nabla ^2}\left( {{\nabla ^2} - {X^2}} \right)}}$, whereas $a = {X^2} + 4{m^2}{{\bf v}^2}$ and $b = 4{m^2}{{\bf v}^2}{A_1} + 2{t_{00}}\frac{{\left( {{A_2} + \alpha } \right)}}{{{A_2}}}$.

Preservation in time of the primary constraint, $\Pi_0$, leads to the usual secondary constraint (Gauss's law) ${\Gamma _1} \equiv {\partial _i}{\Pi ^i} + \frac{m}{2}{\varepsilon ^{ij}}{\partial _i}{A_j} - {J^0} = 0$ and together displays the first-class structure of the theory. It should be further noted that the extended (first-class) Hamiltonian that generates the time evolution of the dynamical variables has the form  $H = H_C
+ \int {d^2 } x\left( {c_0 \left( x \right)\Pi _0 \left( x \right) + c_1
\left( x\right)\Gamma _1 \left( x \right)} \right)$, where $c_0 \left( x
\right)$ and $c_1 \left( x \right)$ are arbitrary functions of space and time.
It is also important to observe that $\Pi^0 = 0$ for all time 
and $\dot{A}_0 \left( x \right)= \left[ {A_0\left( x \right),H} \right] = c_0 \left( x \right)$, which
is completely arbitrary. Hence we discard $A^0 $ and $\Pi^0$. In other words, it is redundant to retain the term containing $A_0$ because it can be absorbed by redefining the function 
$c_1 (x)$. We can, therefore, write  
\begin{eqnarray}
{H} \mspace{-6mu}&=&\mspace{-8mu} \int {{d^2}x} \left\{ { c(x)\left( {{\partial _i}{\Pi ^i} + \frac{m}{2}{\varepsilon ^{ij}}{\partial _i}{A_j} - {J^0}} \right)} \right\} \nonumber\\
 &+&\mspace{-8mu} \int {{d^2}x} \left\{ {\frac{1}{2}{E_i}\Lambda {D_{ij}}{E_j} + \frac{{2m}}{{{A_2}}}\left( {{\bf v} \cdot {\bf s}} \right){E^i}{O^\prime}{\partial _i}B} \right\}\nonumber\\
 &+&\mspace{-8mu} \int {{d^2}x} \left\{ {\frac{1}{2}B\left( {1 - \frac{{4{m^2}{v^2}}}{{{\nabla ^2}}} - 2{O^{\prime \prime}}} \right)B}\right\}, \label{susyBI-25}
\end{eqnarray}
where $c(x) = c_1 (x) - A_0 (x)$. 

Since there is one first class constraint ${\Gamma _1}(x)$ (Gauss's law), according to the usual procedure, we impose a gauge condition such that the full set of constraints becomes second class. A convenient choice is \cite{Gaete:1997eg}
 \begin{equation}
\Gamma _2 \left( x \right) \equiv \int\limits_{C_{\zeta x} } {dz^\nu }
A_\nu\left( z \right) \equiv \int\limits_0^1 {d\lambda x^i } A_i \left( {
\lambda x } \right) = 0,  \label{susyBI-30}
\end{equation}
where $\lambda$ $(0\leq \lambda\leq1)$ is the parameter describing the
space-like straight path $x^i = \zeta ^i + \lambda \left( {x - \zeta}
\right)^i $ , and $\zeta $ is a fixed point (reference point). There is no
essential loss of generality if we restrict our considerations to $\zeta^i=0 
$. We thus obtain the only non-vanishing equal-time Dirac bracket 
for the canonical variables
\begin{eqnarray}
\left\{ {A_i \left( {\bf x} \right),\Pi ^j \left( {\bf y} \right)} \right\}^ * \mspace{-6mu}&=&\mspace{-6mu}\delta{\
_i^j} \delta ^{\left( 2 \right)} \left( {{\bf x} - {\bf y}} \right) \nonumber\\
&-&\mspace{-6mu} \partial _i^x
\int\limits_0^1 {d\lambda x^j } \delta ^{\left( 2 \right)} \left( {\lambda
{\bf x}- {\bf y}} \right).  \label{susyBI-35}
\end{eqnarray}

Making use of this last equation, we can rewrite the Dirac brackets in terms of the magnetic field 
\begin{equation}
B = {\varepsilon _{ij}}{\partial ^i}{A^j}, \label{susyBI-35a}
 \end{equation}
and electric field
\begin{eqnarray}
{E_i}\mspace{-6mu}&= &\mspace{-6mu}{\Lambda ^{ - 1}}\left( {{\delta _{ij}} + \frac{{{\partial _i}{\partial _j}}}{{\left( {{\gamma ^2}\Lambda  - {\nabla ^2}} \right)}}} \right) \nonumber\\
&\times&\mspace{-6mu}\left( {{\Pi _j} + \frac{{4m}}{{{A_2}}}\left( {{\bf v} \cdot {\bf s}} \right){\varepsilon _{kl}}{\partial ^k}{\partial _j}{A^l} - \frac{m}{2}{\varepsilon _{jk}}{A^k}} \right). \label{susyBI-35b}
\end{eqnarray}

We thus find
\begin{eqnarray}
{\left\{ {{E_i}\left( {\bf x} \right),{E_r}\left( {\bf y} \right)} \right\}^ * } \mspace{-9mu}&=&\mspace{-9mu} \frac{{2m}}{{{A_2}}}{\Lambda ^{ - 2}}\left( {{\bf v} \cdot {\bf s}} \right) {{\cal O}^{\prime}} \left( {{\varepsilon _{kr}}{\partial ^k}{\partial _i} - {\varepsilon _{pi}}{\partial ^p}{\partial _r}} \right) \nonumber\\
&\times&\left( {1 + \frac{{{\nabla ^2}}}{\Omega }} \right){\delta ^{\left( 2 \right)}}\left( {{\bf x} - {\bf y}} \right) \nonumber\\
&+& m{\Lambda ^{ - 2}}D_{ij}^{ - 1}D_{rn}^{ - 1}{\varepsilon _{nj}}{\delta ^{\left( 2 \right)}}\left( {{\bf x} - {\bf y}} \right), \label{SusyBI40}
\end{eqnarray}
where $D_{ij}^{ - 1} = {\delta _{ij}} + \frac{{{\partial _i}{\partial _j}}}{{\left( {\Lambda {\gamma ^2} - {\nabla ^2}} \right)}}$, $ 1 + \frac{{{\nabla ^2}}}{\Omega } = 1 + 2\beta {t_{00}}\frac{{{\nabla ^2}}}{{{A_2}{\gamma ^2}\left( {{\nabla ^2} - {X^2}} \right) - 2\beta {t_{00}}{\nabla ^2}}}$ and ${\gamma ^2} = \frac{{{A_2}}}{{2\beta {t_{00}}}}\left( {{\nabla ^2} - {X^2}} \right)$. \\

Also, it may be stated that
\begin{equation}
{\left\{ {B\left( {\bf x} \right),B\left( {\bf y} \right)} \right\}^ * } = 0, \label{SusyBI45}
\end{equation}
and
\begin{eqnarray}
{\left\{ {{E_i}\left( {\bf x} \right),B\left( {\bf y} \right)} \right\}^ * } =  - {\Lambda ^{ - 1}}{\varepsilon _{ij}}{\partial _j}{\delta ^{\left( 2 \right)}}\left( {{\bf x} - {\bf y}} \right). \label{SusyBI50}
\end{eqnarray}

Making use of the foregoing results, we obtain the following equations of motion for the magnetic and electric fields:
\begin{equation}
\dot B\left( {\bf x} \right) =  - {\varepsilon _{ij}}{\partial _i}{E_j}\left( {\bf x} \right), \label{SusyBI55}
\end{equation}
and
\begin{eqnarray}
{\dot E_i}\left( {\bf x} \right) &=& 
\frac{{2m}}{{{A_2}}}{\Lambda ^{ - 1}}\left( {v \cdot s} \right) {{\cal O}^\prime} \left( {{\varepsilon _{kr}}{\partial _i} - {\varepsilon _{ki}}{\partial _r}} \right)\left( {1 + \frac{{{\nabla ^2}}}{\Omega }} \right) \nonumber\\
&\times&{D_{rb}}{\partial ^k}{E_b}\left( {\bf x} \right) \nonumber\\
&+& {\Lambda ^{ - 1}}{\varepsilon _{ij}}\left( {1 - \frac{{4{m^2}{{\bf v}^2}}}{{{\nabla ^2}}} - 2{{\cal O}^{ \prime  \prime }}} \right){\partial _j}{B}\left( x \right)   \nonumber\\   
&+& \frac{{2m}}{{{A_2}}}{\Lambda ^{ - 1}}{\varepsilon _{ij}}{\partial _j}{\partial _k}{{\cal O}^ \prime }{E_k}\left( {\bf x} \right). \label{SusyBI60}
\end{eqnarray}
It is also straightforward to observe that Gauss's law for the present theory reads
\begin{equation}
{D_{ij}}\Lambda {\partial _i}{E_j} - mB - \frac{{4m}}{{{A_2}}}\left( {{\bf v} \cdot {\bf s}} \right){{\cal O}^ \prime }{\nabla ^2}B = {J^0},\label{SusyBI65}
\end{equation}
where ${D_{ij}} = \left( {{\delta _{ij}} - \frac{{{\partial _i}{\partial _j}}}{{\Lambda {\gamma ^2}}}} \right)$.

Next, it is to be specially noted that by taking into account the assumed conditions of static fields, equations (\ref{SusyBI55}) and (\ref{SusyBI60}) must vanish. We accordingly express the magnetic field in the form 
\begin{equation}
B =  - \frac{{m\left[ {\left( {{\bf v} \cdot {\bf s}} \right) - 1} \right]}}{{{A_2}}}\frac{{\left[ {\left( {1 - {A_2}} \right){\nabla ^2} + {A_1}} \right]}}{{\left( {{\nabla ^2} - {X^2}} \right)\left( {{\nabla ^2} + 4{m^2}{{\bf v}^2}} \right)}}{\partial _i}{E_i}. \label{SusyBI70}
\end{equation}
Inserting equation (\ref{SusyBI70}) into equation (\ref{SusyBI65}), we find that the static electric field can be brought to the form
\begin{eqnarray}
{E_i}({\bf x})\mspace{-8mu} &=&\mspace{-8mu} \frac{1}{{{g_1}}}{\partial _i}\left\{ {\frac{{\left[ {{\nabla ^4} + \left( {4{m^2}{{\bf v}^2}{X^2}} \right){\nabla ^2} - 4{m^2}{{\bf v}^2}{X^2}} \right]}}{{{\nabla ^2}\left[ {{\nabla ^4} + \frac{{{g_2}}}{{{g_1}}}{\nabla ^2} + \frac{{{g_3}}}{{{g_1}}}} \right]}}} \right\} \nonumber\\
&\times&\mspace{-8mu}\left( { - {J^0}} \right). \label{SusyBI75}
\end{eqnarray}
Here we have simplified our notation by setting ${g_1} = 1 + 2{t_{00}} + \frac{{2{t_{00}}\left( {\alpha  - \beta } \right)}}{{{A_2}}}$, ${g_2} = \left( {1 + 2{t_{00}}} \right){X^2} + \left\{ {\left[ {1 - \left( {{\bf v} \cdot {\bf s}} \right)} \right]\frac{{\left( {1 - {A_2}} \right)}}{{{A_2}}} - 4{{\bf v}^2}\left( {1 + 2{t_{00}} + 2{t_{00}}\frac{{\left( {\alpha  - \beta } \right)}}{{{A_2}}}} \right)} \right\}$ and ${g_3} = {m^2}{X^2}\left[ {\left( {{\bf v} \cdot {\bf s}} \right) - 4{{\bf v}^2}\left( {1 + 2{t_{00}}} \right) - 1} \right]$. By a further manipulation of the terms, we can write equation (\ref{SusyBI75}) also as    
\begin{eqnarray}
{E_i}\left( {\bf x} \right) \mspace{-6mu} &=& -\mspace{-9mu} \frac{1}{{{g_1}}}\frac{1}{{\left( {M_1^2 - M_2^2} \right)}}{\partial _i}\left[ {\frac{{{\nabla ^2}}}{{\left( {{\nabla ^2} - M_1^2} \right)}} - \frac{{{\nabla ^2}}}{{\left( {{\nabla ^2} - M_2^2} \right)}}} \right] \nonumber\\
&\times& \mspace{-9mu} \left( { - {J^0}} \right)   \nonumber\\
 &+&\mspace{-9mu}\frac{1}{{{g_1}}}\frac{{\left( {{X^2} - 4{m^2}{{\bf v}^2}} \right)}}{{\left( {M_1^2 - M_2^2} \right)}}{\partial _i}\left[ {\frac{1}{{\left( {{\nabla ^2} - M_1^2} \right)}} - \frac{1}{{\left( {{\nabla ^2} - M_2^2} \right)}}} \right] \nonumber\\
&\times& \mspace{-9mu}\left( { - {J^0}} \right)  \nonumber\\
&+&\mspace{-9mu}\frac{{\left( {4{m^2}{v^2}{X^2}} \right)}}{{{g_1}\left( {M_1^2 - M_2^2} \right)}}\frac{{{\partial _i}}}{{{\nabla ^2}}}\left[ {\frac{1}{{\left( {{\nabla ^2} - M_1^2} \right)}} - \frac{1}{{\left( {{\nabla ^2} - M_2^2} \right)}}} \right] \nonumber\\
&\times&\mspace{-9mu}\left( { - {J^0}} \right), \label{SusyBI80} 
 \end{eqnarray}
where $M_1^2 =  - \frac{1}{2}\frac{{{g_2}}}{{{g_1}}} + \frac{1}{2}\sqrt {\frac{{g_2^2}}{{g_1^2}} - 4\frac{{{g_3}}}{{{g_1}}}}$ and $M_2^2 =  - \frac{1}{2}\frac{{{g_2}}}{{{g_1}}} - \frac{1}{2}\sqrt {\frac{{g_2^2}}{{g_1^2}} - 4\frac{{{g_3}}}{{{g_1}}}}$.
Considering, for ${J^0}\left( {\bf x} \right) = q{\delta ^{\left( 2 \right)}}\left( {\bf x} \right)$, expression (\ref{SusyBI80}), becomes
\begin{eqnarray}
{E_i}\left( {\bf x} \right) &=&  - \frac{q}{{{g_1}}}\frac{1}{{\left( {M_1^2 - M_2^2} \right)}}{\partial _i}\left\{ {{\nabla ^2}{G_1}\left( {\bf x} \right) - {\nabla ^2}{G_2}\left( {\bf x} \right)} \right\} \nonumber\\
&+& \frac{q}{{{g_1}}}\frac{{\left( {{X^2} - 4{m^2}{{\bf v}^2}} \right)}}{{\left( {M_1^2 - M_2^2} \right)}}{\partial _i}\left\{ {{G_1}\left( {\bf x} \right) - {G_2}\left( {\bf x} \right)} \right\}    \nonumber\\
&+&\frac{q}{{{g_1}}}\frac{{\left( {4{m^2}{{\bf v}^2}{X^2}} \right)}}{{\left( {M_1^2 - M_2^2} \right)}}{\partial _i}\left\{ {\frac{{{G_1}\left( {\bf x} \right)}}{{{\nabla ^2}}} - \frac{{{G_1}\left( {\bf x} \right)}}{{{\nabla ^2}}}} \right\},\label{SusyBI85}
\end{eqnarray}
where ${G_1}\left( {\bf x} \right) =  - \frac{{{\delta ^{\left( 2 \right)}}\left( {\bf x} \right)}}{{{\nabla ^2} - M_1^2}} = \frac{1}{{2\pi }}{K_0}\left( {{M_1}|{\bf x}|} \right)$ and ${G_2}\left( {\bf x} \right) =  - \frac{{{\delta ^{\left( 2 \right)}}\left( {\bf x} \right)}}{{{\nabla ^2} - M_2^2}} = \frac{1}{{2\pi }}{K_0}\left( {{M_2}|{\bf x}|} \right)$.

With the foregoing information, we can now proceed to obtain the energy interaction. As already mentioned, in order to accomplish this purpose we shall use the gauge-invariant, but path-dependent, variables formalism \cite{Gaete:1997eg}
\begin{equation}
V \equiv q\left( {{{\cal A}_0}\left( {\bf 0} \right) - {{\cal A}_0}\left( {\bf y} \right)} \right), \label{SusyBI90}
\end{equation}
where the physical scalar potential is given by
\begin{equation}
{{\cal A}_0}\left( {\bf x} \right) = \int_0^1 {d\lambda } {x^i}{E_i}\left( {\lambda {\bf x}} \right), \label{SusyBI95}
\end{equation}
and $i=1,2$. As was shown in \cite{Gaete:1997eg}, this follows from the vector gauge-invariant field expression 
\begin{equation}
{{\cal A}_\mu }\left( x \right) \equiv {A_\mu }\left( x \right) + {\partial _\mu }\left( { - \int_\xi ^x {d{z^\mu }{A_\mu }\left( z \right)} } \right), \label{SusyBI100}
\end{equation}
where the line integral is along a space-like path from $\xi$ to $x$, on a fixed time slice. It may be noted that these variables (\ref{SusyBI100}) commute with the sole first class constraint (Gauss's law). From this it follows that these variables are physical variables.

With the aid of equation (\ref{SusyBI85}), equation (\ref{SusyBI95}) becomes
\begin{eqnarray}
{{\cal A}_0}({\bf x}) &=& - \frac{q}{{{g_1}}}\frac{1}{{\left( {M_1^2 - M_2^2} \right)}}\left( {{\nabla ^2}{G_1}\left( {\bf x} \right) - {\nabla ^2}{G_2}\left( {\bf x} \right)} \right) \nonumber\\
&+& \frac{q}{{{g_1}}}\frac{{\left( {{X^2} - 4{m^2}{{\bf v}^2}} \right)}}{{\left( {M_1^2 - M_2^2} \right)}}\left( {{G_1}\left( {\bf x} \right) - {G_2}\left( {\bf x} \right)} \right) \nonumber\\
&+& \frac{q}{{{g_1}}}\frac{{\left( {4{m^2}{{\bf v}^2}{X^2}} \right)}}{{\left( {M_1^2 - M_2^2} \right)}}\left( {\frac{{{G_1}\left( {\bf x} \right)}}{{{\nabla ^2}}} - \frac{{{G_2}\left( {\bf x} \right)}}{{{\nabla ^2}}}} \right), \label{SusyBI105}
\end{eqnarray}
after subtracting the self-energy terms.

We accordingly express the potential for two opposite charges located at ${\bf 0}$ and ${\bf y}$ in the form
\begin{eqnarray}
V &=&  - \frac{{{q^2}}}{{2\pi {g_1}}}\frac{{\left( {{X^2} - 4{m^2}{{\bf v}^2}} \right)}}{{\left( {M_1^2 - M_2^2} \right)}}\left( {{K_0}\left( {{M_1}L} \right) - {K_0}\left( {{M_2}L} \right)} \right) \nonumber\\
&+&\frac{{{q^2}}}{{{g_1}}}\frac{{{m^2}{{\bf v}^2}{X^2}}}{{{M_1}{M_2}\left( {{M_1} + {M_2}} \right)}}L    \nonumber\\
&+& \frac{{{q^2}}}{{2\pi {g_1}}}\frac{1}{{\left( {M_1^2 - M_2^2} \right)}}\left( {{\nabla ^2}{K_0}\left( {{M_1}L} \right) - {\nabla ^2}{K_0}\left( {{M_2}L} \right)} \right),  \nonumber\\
\label{SusyBI110}
\end{eqnarray}
where $L \equiv  |{\bf y}|$. In this last line, we have used that $\frac{{{G_1}\left( {\bf x} \right)}}{{{\nabla ^2}}} = \frac{{|{\bf x}|}}{{4{M_1}}}$ and $\frac{{{G_2}\left( {\bf x} \right)}}{{{\nabla ^2}}} = \frac{{|{\bf x}|}}{{4{M_2}}}$.

\section{Concluding remarks}

The goals of this contribution were two-fold: 

(a) to write down the LSV fermonic matter Lagrangian of fermonic eq.(5) in a supersymmetric scenario along the lines followed in the series of papers cast in Ref. \cite{Belich:2013rma,Belich:2015qxa,HelayelNeto:2010zz},
and

(b) to inspect how the SUSY LSV terms worked out in $(1+3)$-dimensions may go down, by means of dimensional reduction, and affect the interparticle potential of planar Electrodynamics.

We thus find that the three terms on the right hand side of expression (\ref{SusyBI110}) display a screening part, encoded in the Bessel functions and their derivatives, and the linear confining potential. We readily verify that the linear potential disappears when $m, {X^{2}}, or \ {\bf v} \to 0$. Mention should be made, at this point, to the fact that the two first terms on the right hand side of expression (\ref{SusyBI110}) is exactly the result obtained for $D=3$ models of antisymmetric tensor fields that emerges from the condensation of topological defects, as a consequence of the Julia-Thoulousse mechanism \cite{Gaete:2005ht}. This same potential profile is obtained in the case of condensation of charged scalars in $D=3$ dimensions \cite{Gaete:2009ev}. 

A point that we intend next to inspect regards the spin-orbit interaction in two-dimensional materials in presence of SUSY and an external anisotropy as it appears in the Lagrangian of Eq.(\ref{susy3d105b}) above. We shall soon report on this issue elsewhere.

\section{ACKNOWLEDGMENTS}

One of us (P. G.) was partially supported by Fondecyt (Chile) grant 1180178 and by Proyecto Basal FB0821. Y. M. P. G. and L. D. B. express their gratitude to FAPERJ and CNPq-Brazil for their Fellowships. 

\section{Appendix: Conventions and notation for the dimensional reduction}

Let us start our analysis by introducing the way in which the dimensional reduction process will be implemented. Hat indexes are used in $(3+1)$- D, $\hat{\mu} = 0,1,2,3$, whereas normal indexes are used for $(2+1)$- D case, $\mu = 0,1,2$. Also, we mention that the z-component of all vector field $V$ will be represented by a scalar field, i.e., $V^{\hat{\mu}} = (V^\mu, V^3=\zeta)$. Besides, the Dirac matrices will be rewritten in the form:
\begin{equation}
\Gamma^{\hat{\mu}} = \begin{pmatrix} 
\gamma^\mu & 0\\ 
0 & -\gamma^\mu
\end{pmatrix} ~;~ \hat{\mu} = 0,1,2
\end{equation}
\begin{equation}
\Gamma^{\hat{\mu}} = \begin{pmatrix} 
0 & i\\ 
i & 0
\end{pmatrix}~;~ \hat{\mu}= 3
\end{equation}

\begin{equation}
\Gamma_5 = \begin{pmatrix} 
0 & i\\ 
-i & 0
\end{pmatrix},
\end{equation}
\begin{equation}
\Gamma_{R/L} = \frac{1}{2}\begin{pmatrix} 
1 & \pm i\\ 
\pm i & 1
\end{pmatrix}.
\end{equation}
It should be noted that is implicit the 2x2 identity matrix inside the matrices, and $\gamma^0 = \sigma_y$ , $\gamma^1 = \sigma_x$,$\gamma^2 = i \sigma_z$. Let us also mention here that the Dirac spinor $\Psi$ will be splitted into two components, $\Psi = \begin{pmatrix}\Psi_1 ~ \Psi_2 \end{pmatrix}^T$.


\begin{thebibliography}{}
\bibitem{KPRL89} V. A. Kostelecky and S. Samuel, \textit{Phys. Rev. Lett}. 
\textbf{63}, 224 (1989). 

\bibitem{KosteleckyPRL91} V. A. Kostelecky and S. Samuel, \textit{Phys. Rev. Lett}. \textbf{66}, 1811 (1991).

\bibitem{Kostelecky89} V. A. Kostelecky and S. Samuel, \textit{Phys. Rev. }D\textbf{\ 39}, 683 (1989).

\bibitem{KosteleckyPRD89} V. A. Kostelecky and S. Samuel, \textit{Phys. Rev. }D\textbf{\ 40}, 1886 (1989). 

\bibitem{Potting91} V. A. Kostelecky and R. Potting, \textit{Nucl. Phys.} B 
\textbf{\ 359}, 545 (1991). 

\bibitem{Potting96} V. A. Kostelecky and R. Potting, Phys. Lett. B \textbf{381}, 89 (1996). 

\bibitem{Potting95} V.A.Kostelecky and R. Potting, \textit{Phys. Rev. }D \textbf{\ 51}, 3923 (1995).

\bibitem{DColladay97} D. Colladay and V. A. Kosteleck\'{y}, \textit{Phys. Rev.}
D \textbf{55}, 6760 (1997).

\bibitem{Colladay98} D. Colladay and V. A. Kosteleck\'{y}, \textit{Phys. Rev. }D \textbf{58}, 116002 (1998).

\bibitem{Russell11} V. A. Kosteleck\'{y} and N. Russell, \textquotedblleft Data
Tables for Lorentz and CPT Violation,\textquotedblright\ Rev. Mod. Phys. 
\textbf{83}, 11 (2011).

\bibitem{Bailey06} Q.G. Bailey and V.A. Kosteleck\'{y}, Phys. Rev. D \textbf{
74}, 045001 (2006).

\bibitem{Chkareuli} J.L. Chkareuli, EPJ C\textbf {74} (2014) 2906.

\bibitem{Pospelov} M. Pospelov and C. Tamarit, JHEP 01 (2014) 048.

\bibitem{Bailey04} Q. G. Bailey, V.A. Kosteleck\'{y}, Phys. Rev. D \textbf{70} (2004)
076006.

\bibitem{Betschart} G. Betschart, E. Kant, and F. R. Klinkhamer, \textit{Nucl. Phys.} B, \textbf{\ 815}, 198-214 (2009).

\bibitem{Tasson09} V.A. Kosteleck\'{y} and J.D. Tasson, Phys. Rev. Lett. 
\textbf{102}, 010402 (2009).

\bibitem{Lane} V.A. Kostelecky and C. D. Lane, \textit{J. Math. Phys.} \textbf{40}, 6245
(1999).

\bibitem{Lehnert} R. Lehnert, \textit{J. Math. Phys.} \textbf{45}, 3399 (2004).

\bibitem{Mund:2019nap}  J.~Mund, K.~H.~Rehren and B.~Schroer, arXiv:1906.09596 [hep-th].

\bibitem{MacD} D.~Colladay and P.~McDonald, Phys.\ Rev.\ D {\bf 83}, 025021 (2011).
    
\bibitem{Lehum} A.~C.~Lehum, J.~R.~Nascimento, A.~Y.~Petrov and A.~J.~da Silva, Phys.\ Rev.\ D {\bf 88}, 045022 (2013).

\bibitem{Belich:2013rma} H.~Belich, L.~D.~Bernald, P.~Gaete and J.~A.~Helay\"el-Neto,
Eur.\ Phys.\ J.\ C {\bf 73}, no. 11, 2632 (2013).

\bibitem{Belich:2015qxa} H.~Belich, L.~D.~Bernald, P.~Gaete, J.~A.~Helay\"el-Neto and F.~J.~L.~Leal, Eur.\ Phys.\ J.\ C {\bf 75}, no. 6, 291 (2015).

\bibitem{Yu:2019opk} J.~Yu, R.~Roiban, S.~K.~Jian and C.~X.~Liu, Phys.\ Rev.\ B {\bf 100}, no. 7, 075153 (2019).

\bibitem{Gao:2017bqf} P.~Gao and H.~Liu, JHEP {\bf 1801}, 040 (2018).

\bibitem{Rahmani:2015qpa} A.~Rahmani, X.~Zhu, M.~Franz and I.~Affleck, Phys.\ Rev.\ Lett.\  {\bf 115}, no. 16, 166401 (2015). 

\bibitem{Grover:2013rc} T.~Grover, D.~N.~Sheng and A.~Vishwanath, Science {\bf 344}, no. 6181, 280 (2014). 
    
\bibitem{Gaete:2018nwq} P.~Gaete, J.~A.~Helay\"{e}l-Neto and L.~P.~R.~Ospedal,
EPL {\bf 125}, no. 5, 51001 (2019).
 
\bibitem{Gaete:2016tcn} P.~Gaete and J.~A.~Helay\"{e}l-Neto, Adv.\ High Energy Phys.\  {\bf 2016}, 6043548 (2016).

\bibitem{Gaete:2005ht} P.~Gaete and C.~Wotzasek, Phys.\ Lett.\ B {\bf 625}, 365 (2005).

\bibitem{Gaete:2009ev} P.~Gaete and J.~A.~Helay\"{e}l-Neto, Phys.\ ett.\ B {\bf 683}, 211 (2010).

\bibitem{Ferreira:2004hx} M.~M.~Ferreira, Jr., Phys.\ Rev.\ D {\bf 70}, 045013 (2004).

\bibitem{Ferreira:2004ax} M.~M.~Ferreira, Jr., Phys.\ Rev.\ D {\bf 71}, 045003 (2005).

\bibitem{Casana:2011vh} R.~Casana, E.~S.~Carvalho and M.~M.~Ferreira, Jr, Phys.\ Rev.\ D {\bf 84}, 045008 (2011).

\bibitem{Casana:2011du} R.~Casana, M.~M.~Ferreira, Jr and R.~P.~M.~Moreira, Phys.\ Rev.\ D {\bf 84}, 125014 (2011).

\bibitem{Casana:2011fe} R.~Casana, M.~M.~Ferreira, Jr and R.~P.~M.~Moreira, Eur.\ Phys.\ J.\ C {\bf 72}, 2070 (2012).

\bibitem{Ferreira:2019jbx} M. M. Ferreira, J. A.A.S. Reis, M. Schreck, Phys.\ Rev.\ D {\bf 100},  095026 (2019).

\bibitem{Casana:2015kna} R.~Casana, M.~M.~Ferreira, V.~E.~Mouchrek-Santos and E.~O.~Silva,
Phys.\ Lett.\ B {\bf 746}, 171 (2015).

\bibitem{Zarro} J. R. Nascimento, A. Y. Petrov, C. Wotzasek and C. A. D. Zarro, Phys.\ Rev.\ D {\bf 89}, 065030 (2014).

\bibitem{Petrov} M. Gomes, J. R. Nascimento, A. Y. Petrov and A. J. da Silva, Phys.\ Rev.\ D {\bf 81}, 045018 (2010).

\bibitem{Farias} C. F. Farias, A. C. Lehum, J. R. Nascimento and A. Y. Petrov, Phys.\ Rev.\ D {\bf 86}, 065035 (2012).

\bibitem{Silva} A. C. Lehum, J. R. Nascimento, A. Y. Petrov and A. J. da Silva, Phys.\ Rev.\ D {\bf 88}, 045022 (2013). 

\bibitem{Scherk} J. Scherk and J. H. Schwarz, Phys. Lett. B {\bf 57}, 464 (1975).

\bibitem{Gaete:1997eg} P.~Gaete,  Z.\ Phys.\ C {\bf 76}, 355 (1997).

\bibitem{HelayelNeto:2010zz} J.~A.~Helay\"{e}l-Neto, H.~Belich, G.~S.~Dias, F.~J.~L.~Leal and W.~Spalenza,
PoS ICFI {\bf 2010}, 032 (2010).
\end{thebibliography}
\end{document}